\documentstyle[preprint,aps]{revtex}

\begin{document}
 \draft
 \title{Ferroelectric polarization flop in a frustrated magnet MnWO$_4$ induced by magnetic fields
}
 \author{K.\ Taniguchi$^{1}$, N.\ Abe$^{2}$, T.\ Takenobu$^{3}$, Y.\ Iwasa$^{3}$, and T.\ Arima$^{1}$}
\address{
$^{1}$ Institute of Multidisciplinary Research for Advanced Materials, Tohoku University, Sendai 980-8577, Japan
}
\address{
$^{2}$ Department of Physics, Tohoku University, Sendai 980-8578, Japan
}
\address{
$^{3}$ Institute for Material Research, Tohoku University, Sendai 980-8577, Japan
}

 \date{received\hspace*{3cm}}
 \maketitle

 \begin{abstract}
 The relationship between magnetic order and ferroelectric properties has been investigated for MnWO$_4$ with long-wavelength magnetic structure. Spontaneous electric polarization is observed in an elliptical spiral spin phase. The magnetic-field dependence of electric polarization indicates that the noncollinear spin configuration plays a key role for the appearance of ferroelectric phase. An electric polarization flop from the $b$ direction to the $a$ direction has been observed when a magnetic field above 10T is applied along the $b$ axis. This result demonstrates that an electric polarization flop can be induced by a magnetic field in a simple system without rare-earth 4$f$-moments.
 \end{abstract}

 \bigskip
 \narrowtext
   In the past several years, the coupling between ferroelectric (FE) and magnetic order in multiferroics, where the both order coexist, has been attracting much attention \cite{Fiebig}. In particular, a new type of multiferroics such as rare-earth perovskite $R$MnO$_3$ ($R$=Gd,Tb,Dy), in which FE order appears simultaneously at a magnetic transition, has shown a strong interplay between electric polarization and magnetic order \cite{Kimura03,Goto04}. Intriguing phenomena have been found in $R$MnO$_3$ ($R$=Tb,Dy) such as magnetic-field ($H$)-induced electric polarization ($P$) flop \cite{Kimura03,Goto04}. In the recent studies, some other magnetically frustrated systems are also identified to show similar gigantic magnetoelectric (ME) effects \cite{Hur,Kimura05,Lawes,Kimura06,Yamasaki06}. It has been pointed out that there is commonly a close correlation between the long-wavelength magnetic structure and FE order. In particular, the multiferroic materials with long-wavelength magnetic structures appear to be characterized by two key features, which are commensurability and spin chirality (noncollinearity). On the one hand, in the systems such as $R$Mn$_2$O$_5$ ($R$=rare-earth and Y), it has been reported that the locking of the modulation wavelength seems to play a key role for the appearance of FE phase \cite{Chapon04,Chapon06,Kimura06con}. On the other hands, the systems such as rare-earth perovskite $R$MnO$_3$ ($R$=Gd,Tb,Dy) are regarded as another type of multiferroic materials, in which the important factor to cause FE transition is a noncollinear spin configuration, but not the change of the modulation wavelength. In fact, it has been confirmed by the neutron diffraction experiments in perovskite $R$MnO$_3$ ($R$=Tb,Tb$_{1-x}$Dy$_x$), that the inversion symmetry breaking occurs in transverse spiral spin phase without changing the modulation wavelength \cite{Kenzelmann,Arima}. 

As for the mechanism of the FE phase transition at zero magnetic-field, the recent investigations are identifying the dominant parameters. However, the effect of magnetic field has not been fully understood. While rare-earth perovskite $R$MnO$_3$ ($R$=Gd-Tb) and $R$Mn$_2$O$_5$ ($R$=rare earth element) are investigated intensively as typical multiferroic systems with long-wavelength magnetic structures, their responses to magnetic fields are very complex because of the existence of the rare-earth $f$-electron moments and crystal structural complexity \cite{Goto05}. For the sufficient understanding of $H$-induced gigantic ME effects, it is necessary to investigate multiferroic compounds with only one kind of magnetic ions and a simple ion-network. However, the $H$-induced sudden $P$ flop was only observed in perovskite $R$MnO$_3$ with large 4$f$ moments so far to our knowledge. Hence, the 4$f$-3$d$ exchange interaction might be speculated to be one of the key factors for the $P$ flop. In this paper, we report that MnWO$_4$ is multiferroic material, which shows ferroelectric $P$ flop by the application of a magnetic field. MnWO$_4$ is a system with long-wavelength magnetic structure \cite{Lautenschlager}, and contains only one kind of magnetic ions, Mn$^{2+}$. In this material, it has been found that the spiral spin phase (AF2-phase) plays a dominant role for appearance of the FE phase. In addition, we demonstrate that the direction of the FE polarization can be flopped from the $b$ to the $a$ axis by the application of a magnetic-field. Since MnWO$_4$ contains no rare-earth element, the present results indicate that the $H$-induced $P$ flop can take place without help from 4$f$ moments. 

MnWO$_4$ is crystallized in wolframite structure, which belongs to the monoclinic space group $P2/c$ with $\beta$$\sim$91$^{\circ}$ at room temperature (see Fig.\ \ref{fig:Structure}). The crystal structure is characterized by alternative stacking of manganese- and tungsten-layers parallel to the (100) plane (Fig.\ \ref{fig:Structure}(b)). As shown in Fig.\ \ref{fig:Structure}(a), Mn$^{2+}$ ions (S=5/2) are surrounded by distorted oxygen octahedra and aligned in zigzag chains along the $c$ axis. Although this structure may be regarded as one-dimensional Heisenberg-spin chain, it is known that three-dimensional long-range antiferromagnetic (AFM) order is realized since the influence of the spins, which are further than the nearest neighbor one, cannot be neglected \cite{Ehrenberg99}. While other wolframite $M$WO$_4$ ($M$=Fe,Co,Ni) show only one magnetic transition to the commensurate magnetic (CM) state with a propagation vector $\mbox{\boldmath $k$}$=(1/2,0,0), MnWO$_4$ undergoes successive magnetic phase transitions at $\sim$13.5 K (T$_N$),$\sim$12.7 K (T$_2$),$\sim$7.6 K (T$_1$) related to three long-wavelength magnetic ordering states, AF3, AF2 and AF1 \cite{Lautenschlager}. According to the neutron diffraction results, AF1 (T$<$T$_1$) is a CM-collinear AFM-phase, AF2 (T$_1$$<$T$<$T$_2$) is an incommensurate (ICM) elliptical spiral-phase, and AF3 (T$_2$$<$T$<$T$_N$) is an ICM-collinear AFM-phase. The respective propagation vectors are $\mbox{\boldmath $k$}$=($\pm$1/4,1/2,1/2) for AF1 and $\mbox{\boldmath $k$}$=($-$0.214,1/2,0.457) for AF2 and AF3. In AF1 and AF3, magnetic moments collinearly align in the $ac$-plane forming an angle of about 35$^{\circ}$ with the $a$ axis, whereas in AF2 an additional component in the [010] direction exists, as shown in Fig.\ \ref{fig:Structure}(b), (c) respectively \cite{Lautenschlager}. 

Single crystals of MnWO$_4$ were grown by the floating zone method. The resulting crystals appear to be blood red and transparent in thin section. The crystals were oriented using Laue x-ray photographs, and cut into thin plates with wide faces perpendicular to the crystallographic principal axis $a$ or $b$. Gold electrodes were then sputtered onto the opposite faces of the samples for measurements of dielectric constant $\varepsilon$ and electric polarization $P$. We measured $\varepsilon$ at 1kHz using a LCR meter, and obtained $P$ by integration of pyroelectric current, which was measured with an electrometer. The measurements of $\varepsilon$ and electric polarization $P$ in magnetic field up to 14.5T were performed at High Field Laboratory for Superconducting materials, Institute for Materials Research, Tohoku University, Japan. Magnetic susceptibility was measured by a commercial superconducting quantum interference device magnetometer, applying magnetic fields parallel to each crystal principal axis.

Figure\ \ref{fig:Tempdep}(a) displays the temperature dependence of the magnetic susceptibility ($\chi$) in a magnetic field of 0.1T parallel to the $a$, $b$, and $c$ axis respectively. At N\'{e}el temperature, T$_N$$\sim$13.5 K, a cusp is observed in $\chi$ for all the direction. At higher temperatures than T$_N$, $\chi$(T) follows the Curie-Weiss law with a negative Weiss temperature $\theta$. We estimated from $\chi_a$ that $\vert\theta\vert$ is 78 K and the effective moment $\mu_{eff}$ is 6.0 $\mu_B$, which is close to the value 5.9 $\mu_B$ expected from Mn$^{2+}$ (S=5/2). It should be noticed here that T$_N$ is much smaller than $\vert\theta\vert$ in this compound. From the present study, the ratio of $\vert\theta\vert$ to T$_N$ is estimated to be $\sim$6. This indicates that MnWO$_4$ is a spin system with frustration where the antiferromagnetic exchange interactions with the next-nearest neighbor spins should be considered. At the phase transition temperature from AF3 to AF2, T$_2$, an appreciable anomaly is observed only in $\chi_b$. This would reflect the fact that an additional magnetic component along the [010] direction arises at T$_2$ \cite{Lautenschlager}. At T$_1$, which is the transition temperature from AF2 to AF1, $\chi_b$ shows a steep rise, while $\chi_a$ and $\chi_c$ show a sharp drop. This behavior is consistent with the neutron diffraction result that the easy axis of the Mn$^{2+}$ moments is within the $ac$-plane in AF1 \cite{Lautenschlager}. 

Figures\ \ref{fig:Tempdep}(b) and (c) show temperature dependence of $\varepsilon_b$ and $\Delta P_b$ in zero magnetic field. With decreasing temperature, $\varepsilon_b$ shows a sharp peak at T$_2$ and a very small drop ($\sim$0.08\%), which is displayed in a magnified scale, at T$_1$. As seen in Fig.\ \ref{fig:Tempdep}(c), the spontaneous polarization exists in the AF2-phase between T$_1$ and T$_2$ \cite{Explanation}. We have also confirmed the sign reversal of $\Delta P_b$ in the AF2-phase with cooling process in a negative electric field (not shown). These results evidently indicate that MnWO$_4$ becomes ferroelectric simultaneously when AF2-phase with the spiral spin configuration appears. Comparing MnWO$_4$ with $R$MnO$_3$ ($R$=Tb,Dy) \cite{Kimura03,Goto04}, which also show ferroelectricity in spiral phase, the maximum of $P$$\sim$50$\mu$C/m$^2$ is by more than an order of magnitude smaller. This fact may be related with the difference in orbital momentum L between Mn$^{3+}$ (L=2) and Mn$^{2+}$ (L=0). The latter has no net spin-orbit coupling in the ground state. 

Figures\ \ref{fig:Flop}(a) and (b) present the temperature dependence of the electric polarization, which is parallel to the $b$ and the $a$ axis, respectively, in several magnetic fields applied parallel to the $b$ axis. With increasing the magnetic field, the temperature range, where nonzero $\Delta P_b$ appears, becomes narrower and $\Delta P_b$ is suppressed [Fig.\ \ref{fig:Flop}(a)]. On the other hand, a rapid growth of $\Delta P_a$ is observed above 10T [Fig.\ \ref{fig:Flop}(b)]. In Fig.\ \ref{fig:Flop}(c) and (d), we display the magnetic field dependence of $\Delta P$ in each direction and $\varepsilon_a$ at 10.5K. Around 10.7T, where peak anomaly appears in $\varepsilon_a$, $\Delta P_b$ shows drastic suppression, whereas $\Delta P_a$ increases sharply. This contrast means that the magnetic field switches the ferroelectric polarization direction from the $b$ to the $a$ axis. Although $\Delta P_a$ shows small nonzero value in low magnetic fields below 10T, the observed polarization may be caused by a leak component of $\Delta P_b$ due to a small misalignment of the sample axis, since the temperature range of $\Delta P_a$ completely coincides with that of $\Delta P_b$.

In Figure\ \ref{fig:Phase}, we show the magnetoelectric phase diagram of MnWO$_4$ for the direction of magnetic field ($H$) along the $b$ axis. Closed and open squares represent the data points in the cooling (or $H$ decreasing) and warming (or $H$ increasing) runs of pyroelectric current (or $\varepsilon$) measurements, respectively. Reflecting the first-order-like change in $\chi$ and $\Delta P$, hysteresis is observed at T$_1$. On the other hand, hysteresis is hardly observed at T$_2$. Since $\chi$ and $\Delta P$ show continuous change, the phase transition between AF2 and AF3 would be a second-order one. These results are in a good agreement with the fact that the jump of the magnetic propagation vector is only observed at T$_1$ \cite{Lautenschlager}. According to the previously reported magnetic phase diagram \cite{Ehrenberg97}, the FE phase with $P$//$b$ is identical with the elliptical spiral spin phase, AF2. This fact demonstrates that the noncollinear spin configuration plays a key role in the FE phase of MnWO$_4$. 

Several groups have been proposed the mechanism of magnetically driven ferroelectricity for noncollinear magnets \cite{Lawes,Katsura,Mostovoy}. The microscopic model taking into account the spin-orbit interaction \cite{Katsura} predicts that the $P$ direction should be given by $<$$\mbox{\boldmath $e$}_{ij}\times (\mbox{\boldmath $S$}_i\times \mbox{\boldmath $S$}_j)$$>$, in which $\mbox{\boldmath $e$}_{ij}$ is the unit vector connecting two magnetic ions and $\mbox{\boldmath $S$}_\alpha$ ($\alpha=i,j$) is a magnetic moment at the $\alpha$ site. Since the basal plane of the spiral is reported to be inclined to $ab$-plane about 34$^{\circ}$ \cite{Lautenschlager}, the $\mbox{\boldmath $S$}_i\times \mbox{\boldmath $S$}_j$ lies in the $ac$-plane forming an angle of about 124$^{\circ}$ with the $a$ axis. Taking into account that the Mn$^{2+}$ chain is along the $c$ axis ($<$$\mbox{\boldmath $e$}_{ij}$$>$=[001]), the predicted $P$ direction is along the $b$ axis, which is in a good agreement with the present results (See Fig.\ \ref{fig:Flop}). 
 
As displayed in Fig.\ \ref{fig:Phase}, the new FE phase with $P$ parallel to the $a$ axis has also been found in magnetic fields higher than 10T. Since a jump of magnetization has been reported around the phase boundary in the previous research \cite{Ehrenberg97}, the $P$ flop would be caused by a phase transition to a new magnetic phase, which is different from AF1, AF2, and AF3. The identification of this new magnetic phase is a future issue. 

In summary, we have identified MnWO$_4$ as a multiferroic compound in which magnetic and ferroelectric orders coexist in the same temperature range. The magnetic-field dependence shows the close relation between the stability of ICM-elliptical spiral AF2 and the appearance of spontaneous polarization parallel to the $b$ axis. This fact means that the noncollinear spin configuration plays a key role for the ferroelectricity in MnWO$_4$.  It is also found that the electric polarization flops from the $b$ to the $a$ axis when magnetic field is applied along the $b$ axis. This is the first example of the ferroelectric polarization flop induced by magnetic fields in transition-metal oxide systems without rare-earth 4$f$-moments.

We thank M. Saito and S. Otani for help with experiments. This work was partly supported by Grants-In-Aid for Scientific Research from the MEXT Japan.



 \begin{figure}
\caption{
 (Color online) (a) Crystal structure of MnWO$_4$ viewed along the $a$ axis: each Mn atom (purple) is surrounded by an oxygen (red) octahedron. W atoms (grey) separate zigzag chains of Mn atoms. (b) Collinear magnetic structure in AF1 and AF3: magnetic moments lie in the $ac$-plane and canted to the $a$ axis by about 35$^{\circ}$. (c) Elliptical spiral spin structure in AF2: basal plane of spiral is inclined to the $ab$-plane.
\label{fig:Structure}
}
\end{figure}

 \begin{figure}
\caption{
(Color online) (a) Magnetic susceptibility for each crystal axis as a function of temperature in 0.1T. (b) Dielectric constant for $E$//$b$ in 0T. (c) Electric polarization $\Delta P$//$b$ in 0T. During the measurement of pyroelectric current to obtain $\Delta P$, an electric field of 500kV/m was continuously applied along the $b$ axis in a cooling process. $\Delta P$ was calculated by integrating the measured pyroelectric current with respect to time.
\label{fig:Tempdep}
}
\end{figure}

 \begin{figure}
\caption{
(Color online) (a),(b) Temperature dependence of electric polarization in several magnetic fields. (c) Electric polarization, $\Delta P$ along the $b$ (closed circle) and the $a$ axis (open circle) plotted against magnetic field at 10.5K. Plotted values are deduced from (a) and (b). (d) Change in dielectric constant $\varepsilon$//$a$ with sweeping the magnetic field at 10.5K.
\label{fig:Flop}
}
\end{figure}

 \begin{figure}
\caption{
Magnetoelectric phase diagram of MnWO$_4$ in magnetic fields parallel to the $b$ axis. Closed and open squares represent the data points in the cooling (or $H$ decreasing) and warming (or $H$ increasing) runs of pyroelectric current (or $\varepsilon$) measurements, respectively. Closed triangles show magnetic phase transition temperatures determined by the magnetic susceptibility measurements in a cooling run in the present study. Open circles represent the N\'{e}el temperatures reported by another group [18].
\label{fig:Phase}
}
\end{figure}


\begin{references}
\bibitem{Explanation} Although there is a small non-zero polarization in AF1-phase, it is within experimental error, which arises from the background noise ($\sim$0.5pA) in the pyroelectric current integrating process. Concerning the effect of non-zero electric field effect on the observed $P$ value, we confirmed that the AF2-phase shows almost the same magnitude of $P$ in a field-cooled-zero-field-warning runs as in a field cooling run. Therefore, the influence of an applied electric field for $P$ values seems to be small enough.
\end{references}
 \end{document}